\documentclass{iaus260}
\usepackage{graphicx}
\pubyear{2011}
\volume{260}  
\pagerange{346--353}
\jname{The R\^ole of Astronomy in Society and Culture}
\editors{D. Valls-Gabaud \& A. Boksenberg, eds.}

\title[A palace for astronomy in Buenos Aires]                   
{A palace for astronomy in Buenos Aires} 

\author[Alejandro Gangui]           
{Alejandro Gangui$^{1,2}$}        

\affiliation{$^1$Instituto de Astronom{\'\i}a y F{\'\i}sica del Espacio / CONICET, \\ 
                 Ciudad Universitaria, 1428 Buenos Aires, Argentina. \\ 
                 email: {\tt gangui@df.uba.ar} \\[\affilskip]
             $^2$Centro de Formaci\'on e Investigaci\'on en la Ense\~nanza de las Ciencias, \\ FCEyN, Universidad de Buenos Aires.}

\begin{document}
\maketitle

\begin{abstract}

In no other epoch of Western history like in the Middle Ages, cosmology was so key an element of culture and, one way or another, the
motion of the heavens ended up impregnating the literature of that time. Among the most noteworthy poets we find Dante Alighieri, who
became famous for his {\it Commedia}, a monumental poem written roughly between 1307 and his death in 1321, and which the critics from
XVIth century onwards dubbed {\it Divina}. In this and other works, Dante pictures the cosmic image for the world, summing up the current
trends of Neoplatonic and Islamic traditions.  The Barolo Palace in the city of Buenos Aires is a singular combination of both astronomy
and the worldview displayed in Dante's poetic masterpiece. Some links of the Palace's main architectural structure with the three realms
of the Comedy have been studied in the past. In this note we consider its unique astronomical flavor, an issue which has not been
sufficiently emphasized yet.

\keywords{Medieval astronomy, Dante's cosmology, Eclectic architecture} 
\end{abstract}

\firstsection 
\section{Introduction}

The universe of Dante is described by Christian theology and Ptolemaic astronomy. It is based on a simplified Aristotelian model, where
the sphere of the Earth is standing still at the geometric center of the universe. The northern hemisphere is the only habitable place
where man can dwell, and has Mount Zion at its center near Jerusalem. Ninety degrees to the East, one finds the river Ganges, and ninety
degrees to the West, the Ebro in the Iberian Peninsula. The southern hemisphere has no dry land emerging from the immense ocean covering
it and is therefore forbidden to man. However, a huge mountain lies just at its center, exactly opposite to Mount Zion, and so, both
rivers and both mountains inscribe a cross within the sphere of the Earth (Borges 1982). Below Zion, with the form of an inverted cone, we
find Hell, the place where sinners are punished. It is divided into nine decreasing concentric circles which culminate at the center of
the Earth, and therefore also of the universe. From there, there is a crack, or narrow tunnel, opened up by the waters of the river Lethe,
the river of forgetfulness in Hades, which connects the bottom of Hell with the base of the mountain of Purgatory, in the opposite
hemisphere. This second realm of the afterlife, according to Dante, is thus located in the midst of the unexplored ocean of the
uninhabited hemisphere directly opposite to Jerusalem. And, should the souls reach the top of the mountain, they would be cleansed of all
sin and made perfect, and therefore wait their turn to ascend to Heaven.

Dedicated efforts of many people in Buenos Aires led to a monumental recognition for the topography of Dante's cosmology, which was shaped
in reinforced concrete. The Barolo Palace is this monument, and still today it offers a vivid and prominent example of the role played by
medieval astronomy in culture in modern times.

\section{Palanti's eclectic project}

The Italian architect Mario Palanti (1885-1979) was in charge of the works of the Italian Pavilion for the Argentine Centenary Exhibition
of 1910. After returning to his home country and serving in the First World War, in 1919 he traveled back to Argentina. This time, Palanti
reaches the southern hemisphere with one single idea in mind: to materialize a votive offering, in the guise of a monumental Temple,
honoring Dante in the 600th anniversary of his death. In the past, such temples were made in order to gain favor with supernatural forces,
may be the very same forces that --Palanti imagined-- pushed Dante along his trip though Hell, Purgatory and Paradise.  The temple had to be
located in the ascensional axis of the souls dwelling in these last two realms, in the antipodes of Mount Zion. In Buenos Aires, Palanti met
Luis Barolo and the project took shape.

Barolo (1869-1922) was an Italian businessman and the needed wealthy sponsor for Palanti's dream. Like the architect, he was a lover of
his native country and of Dante, {\it il sommo poeta}. Moreover, like many immigrants, he was worried about international military
conflicts and the necessity of preservation of European culture against the disasters of the war. Having met Palanti at the summit of his
fortune, Barolo agreed to finance Palanti's project aiming to immortalize the Italic culture and Dante's verses.

Palanti wanted his future monumental building to be located in a place diametrically opposite to Mount Zion.  But that antipodal point was
very far away from the capital city of Argentina. Let us recall that Dante and Virgil, after having escaped from the abyss of Hell through
a natural underground passage (the {\it natural burella}), emerged on the shores of an island located in the southern hemisphere, where
they found the mountain of Purgatory. Now, Jerusalem's coordinates are roughly (31$^{\rm o}$ 47' N, 35$^{\rm o}$ 13' E), so that its
antipodes fall approximately at Buenos Aires correct latitude (-34$^{\rm o}$ 36', for Buenos Aires city, to be precise) but missing
completely the longitude of the site (-58$^{\rm o}$ 20' for Buenos Aires). Indeed, the mythic mountain, according to most of all
a-bit-too-realistic topographical representations of the death realms depicted for Dante's journey (Gangui 2009), ought to be located at
longitude -144$^{\rm o}$ 47', which clearly makes the famous mountain fall in the middle of the Pacific Ocean.

But this fact does not get Palanti down. The architect probably recalled Dante's verses ({\it Inferno} XXVI, 85-142) on Ulysses' last
voyage to the southern hemisphere when, endangering his sailors with the aim to gain knowledge of the unknown, the hero from Ithaca
traverses those Pillars ``where Hercules his landmarks set as signals,/ That man no farther onward should adventure'' (Inf. XXVI,
108-109). And so Palanti imagines a new set of fabulous Pillars of Hercules, but this time in the River Plate estuary. Thus, he plans the
construction of two votive monuments, one in each side of the River Plate: the Salvo Palace, to be located in Montevideo, Uruguay, and the
Barolo Palace on the shores of Buenos Aires.

\section{The Barolo palace}

The Palace constructed in Buenos Aires was conceived to be a materialization of the Divine Comedy. As such, it is full of allegories and
symbolisms referring to Dante's epic poem. For example, the building is divided vertically into three parts, exactly as the many pictorial
``representations'' depict Dante's three realms of the afterlife. Palanti designed the building so that it became a scale model of Dante's
universe, following the tradition of the Gothic cathedrals, the supreme Opus of stonework in the Middle Ages (Hilger 1993). The arched
walls of the ground floor passage constructed by the architect join together forming a series of vaults, with a height equivalent to three
standard levels, and which are filled with Latin inscriptions: from Virgil's Eclogues, biblical writings, Ovid, Horace, etc.  As it was
usual of Palanti's eclectic constructions, the palace was conceived as a lay temple for the promotion of liberal arts.

The building itself turned out to be monumental, considering the time when it was erected, of course: it comprises 22 floors and 2
basements, reaching a height of approximately 100 meters. This detail was not by chance: the Divine Comedy comprises exactly 100 cantos,
which for Dante represented the ``perfect number'' (i.e., 10) multiplied by itself (Gangui 2008). The city code of the time, however,
limited the maximum height allowed for new buildings along {\it Avenida de Mayo} in the city of Buenos Aires, where the Palace is located,
and so a special permission had to be granted. The Barolo, as it is now known, set quite a few ``firsts'', being considered the first high
building of the city, and indeed the highest of Latin America of its time. It continued being the highest of Argentina until the
construction of the Kavanagh building in 1936, and it was the first entirely made of reinforced concrete, a technique with which Palanti
had good experience, but which had no precedents in Argentina by that time. 

\begin{figure}[h!tb]
\begin{center} 
\includegraphics[width=13.4cm]{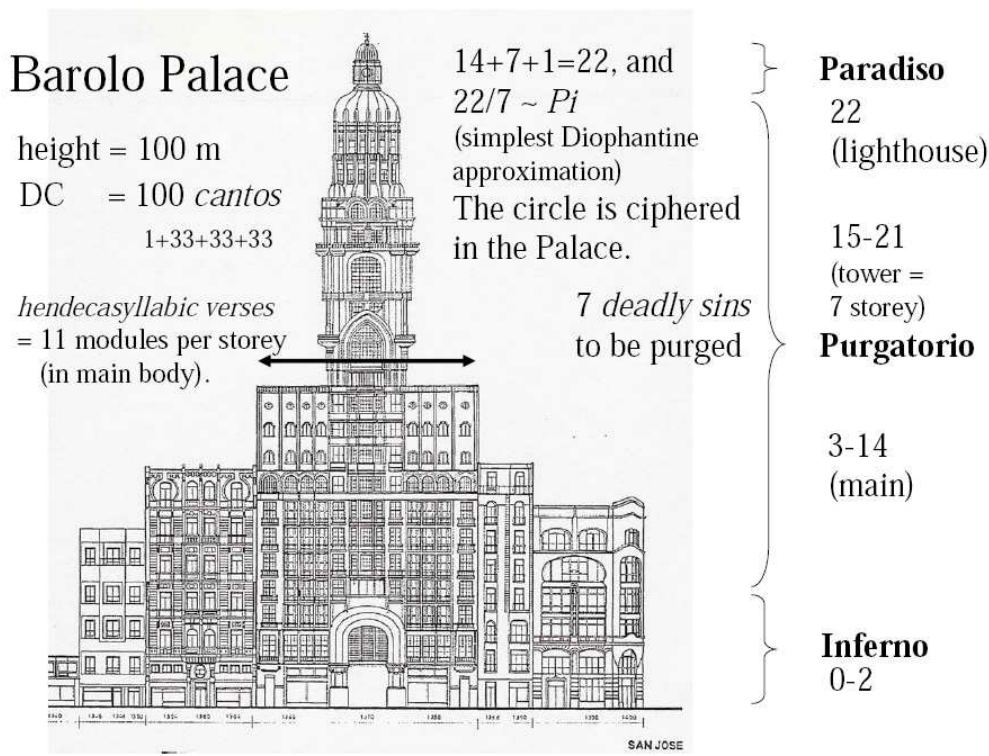}
\caption{Some of the peculiarities of the Barolo Palace designed by architect Palanti.}\label{fig0}\end{center}\end{figure} 

As soon as one leaves the ground level and its vaults (worthy a three-story building, as we said) and takes up the stairs, on the third
floor one enters the cleansing realm of Purgatory. From floor 3 to the 14th, one goes up -one ``terrace'' after the other- as Dante did
along the mountainside. Then, after the 14th, one can climb up more seven stories through the tower of the Palace, a clear allegory of the
seven deadly sins to be purged in each of the terraces of Purgatory. At the very top, the Palace in crowned with a rotating lighthouse,
the material representation of the angelic choruses of Dante's Paradise. This is the house of an arc lamp (voltaic arc lamp) of roughly
300,000 bougies (approximately 300,000 candelas). All these numbers appropriately mixed together (in a sort of numerological portrait of
reality, so appreciated by Pythagoreans, and therefore also by Dante) give us: 14 + 7 + 1 = 22. Now, by dividing this result by 7 (seven
deadly sins or stories through the tower), it yields 22/7 which is the simplest Diophantine approximation of number $\pi$, the ratio of a
circle's perimeter to its diameter. Again, there is no coincidence in this; Palanti twisted his design to make the circle --the
most perfect geometric figure-- ciphered within the Palace's architecture.

Of course, taking this ``realistic'' iconography --and especially Dante's own words-- one cannot rightly say that Purgatory lies ``on top
of'' Hell, for --as we saw above-- these two realms are ``located'' in opposite hemispheres of the Earth. Dante ``the writer'' suggests
this when, having arrived to the center of the Earth and descended along the body of giant Lucifer (the ``fell worm who mines the
world''), he abruptly gets upside down after passing a certain point (Inf. XXXIV, 106-111). Surprised, Dante ``the pilgrim'' asks Virgil
to explain to him what had happened, getting as a reply: ``[...] Thou still imaginest/ Thou art beyond the centre, where I grasped/ The
hair of the fell worm, who mines the world./ That side thou wast, so long as I descended;/ When round I turned me, thou didst pass the
point/ To which things heavy draw from every side''.

Dante and his guide had just gone through that special ``point'' onto which all weights converge in the search for their {\it natural
  place}, in the very sense Aristotle conceived it. Having passed to the southern hemisphere they then stopped descending and began
climbing to reach the surface opposite to Jerusalem: ``We mounted up, he first and I the second,/ Till I beheld through a round aperture/
Some of the beauteous things that Heaven doth bear;/ Thence we came forth to rebehold the stars.'' (Inf. XXXIV, 136-139; recall, all three
canticas end with the word {\it stars}). The pilgrims are now ready to start the second part of their journey in Purgatory.

So, Purgatory strictly speaking is not ``above'' Hell (within Dante's architecture of the world, one can even claim that Lucifer has both
his head and his legs pointing ``upwards''). The physical partition of the Barolo Palace, however, allegorically needs Hell to be below
all the rest, and this first realm of Dante's trip is represented by the ground floor. So Palanti envisioned Hell as the main entrance of
the Palace, as it was the case for Dante when he went astray (allegorically) in the dark wood, at the very entrance of Hell (Inf. I,
1-3). And, like the hollow void of Hell which is divided into nine infernal circles of punishment, also the central passage of the Palace
(the pedestrian alley which connects two parallel streets, on the ground floor level) has nine vaults of access: three towards {\it
  Avenida de Mayo}, which is the main entrance of the Palace; three towards the former {\it Victoria} street (now {\it Hip\'olito
  Yrigoyen} street), parallel to {\it Avenida de Mayo}; one given by the central vault, which stretches out towards the dome; and two
other ones that contain the staircases, on both sides of the central vault.

\section{Barolo's Purgatory}

Let us note that the two staircases, on both sides of the central vault, are symmetric. Hence, one of these makes the passerby go up
--heading for the second part of the building, namely the Purgatory-- turning to the right, while the other makes her turn to the
left. Now, in the {\it Commedia} there is a preferred way for the pilgrims to descend along the circles of Hell, downwards to the center
of the Earth, and then to climb upwards along the terraces of Mount Purgatory (Gangui 2008). 

Usual representations of Dante's cosmos show the location on Earth of two of the three landmarks of the afterlife: Hell and
Purgatory. Originated from a violent and sudden cataclysm, both parts are cone-like and complementary, as the first is a hollow void while
the second is a solid mass. The descent of the pilgrims proceeds along the circles of the concave walls interior to the cavity of Hell,
which spiral down to the left, whereas the ascent along the convex path of the terraces of Mount Purgatory winds up to the right. Given
the inverse placement of both realms, the trajectory of the pilgrims is always in the same rectilinear direction towards the top of Mount
Purgatory and the heavenly spheres, the Empyrean and, finally, towards God (Cornish 1993).

\begin{figure}[h!tb]
\begin{center} 
\includegraphics[width=13.4cm]{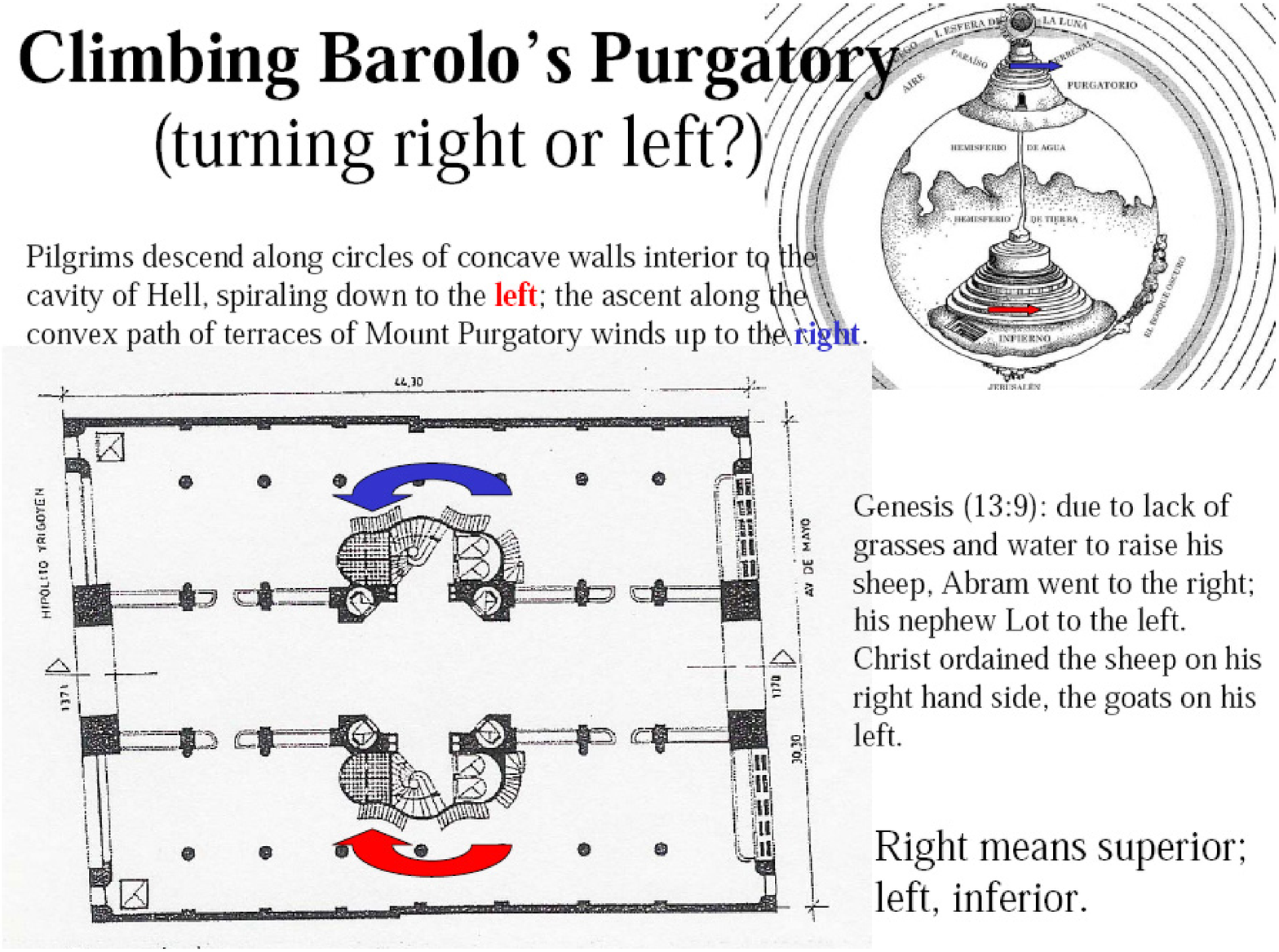}
\caption{Symmetric staircases of The Barolo and the path taken by Dante during his journey.}\label{fig0}\end{center}\end{figure} 

This orientation during their journey is not arbitrary, but has an ancient tradition behind it. Right means superior; left, inferior. In
{\it Genesis} (13:9), due to the lack of grasses and water to raise his sheep, Abraham went to the right and his nephew Lot to the
left. Christ ordained that the sheep should stay at his right hand side, while the goats should remain on his left. In {\it Koran} (sura
69, ayat 13-31), we read ``when the trumpet shall sound one blast [...] On that day ye will be exposed; not a secret of you will be
hidden. [...] as for him who is given his record in his right hand [...] he shall be in a life of pleasure, in a lofty garden'' But, ``as
for him who is given his record in his left hand [...] Take him and fetter him! And burn ye him in the Blazing Fire!'' It does not seem
strange, then, that Dante chose the path on the left when he was traversing Hell, but changed to the path on his right hand side during
his journey along the terraces of Purgatory.

However, it is clear that not both staircases of the Barolo building --as they were built by Palanti, and can still be seen in the detailed
drawing to scale of the construction-- can satisfy the requirement of spiraling up and to the left, for the first two floors, and then up
and to the right, for the rest of the main building. Tourist guides might then suggest visitors to take the appropriate staircase --first,
one of them and, on the third floor, the opposite one-- when the elevators (nine in total, as it should be) are not operating properly.

\section{\it ``...I saw the sign that follows Taurus, and was in it.''}

From the Garden of Eden --the earthly paradise on top of Mount Purgatory-- located just below the Aristotelian sphere of fire, it is
Beatrice who guides Dante along the last leg of his ultra mundane travel, through the nine celestial spheres of Heaven. Let us recall that
Virgil, as a pagan, had forbidden access to Paradise (Inf. IV, 39), and was forever obliged to remain in Limbo, the first circle of
Hell. The Paradise is the abode of angels and saints, and was structured, according to Judeo-Christian tradition, by the classical nine
concentric, spherical and incorruptible, heavens: the eight astronomical Ptolemaic ``spheres'' plus the Aristotelian primum
mobile. Surrounding this all, there was the Empyrean, as the tenth and last theological sphere, infinite and motionless, made just of
light and beyond space and time, as its dwelling was in the mind of God.

One sphere after the other the pilgrims pass from the heaven of the Moon to that of Saturn. Once in this last heaven, Dante and Beatrice
are ready to ``jump'' into the sphere of the stars. It is in fact at this moment that the poet hints to us his probable date of birth. In
Canto XXII, 106-111 of Paradise, Dante speaks directly to the reader and tells her that he saw the constellation following Taurus, namely
Gemini. Then, suddenly, faster than when you put ``thy finger in the fire and drawn it out again'', he found himself inside of this new
constellation.

In the following verses, he sings to ``his'' stars: ``O glorious stars, O light impregnated/ With mighty virtue, from which I acknowledge/
All of my genius, whatsoe'er it be,/ With you was born, and hid himself with you,/ He who is father of all mortal life,/ When first I
tasted of the Tuscan air'' (Par. XXII, 112-117). Few people have ever written to themselves a more poetic and touching ``birth
certificate''. What the poet is telling us with these verses is that, in conjunction with the constellation Gemini, it rose and set that
celestial body who is ``father of all mortal life'' --the Sun-- when Dante breathed the air of Tuscany for the first time, namely at his
birthday. From his own words, then, one can narrow Dante's birth date to between May 22 and June 21, 1965, approximately.

So, Gemini seems to be Dante's own constellation and birth sign, and it is from it that he afterwards carries on with his trip through the
heavens. In a sense, we see here that Dante enters the higher spheres of Paradise through this ``door''. However, common wisdom surrounding
the Barolo history takes the Southern Cross as Dante's ``gateway to the heavens'' (Hilger 1996). Why?

There is indeed a famous passage in Purgatorio related to some ``four stars'' which not few commentators relate to the constellation
Crux. Let us read Dante's verses; we will comment on them below: 
``The beauteous planet, that to love incites,/ Was making all the orient to laugh,/ Veiling the Fishes that were in her escort./ To the
right hand I turned, and fixed my mind/ Upon the other pole, and saw four stars/ Ne'er seen before save by the primal people./ Rejoicing
in their flamelets seemed the heaven./ O thou septentrional and widowed site,/ Because thou art deprived of seeing these!/ When from
regarding them I had withdrawn,/ Turning a little to the other pole,/ There where the Wain had disappeared already'' (Pur. I, 19-30).

\begin{figure}[h!tb]
\begin{center} 
\includegraphics[width=13.4cm]{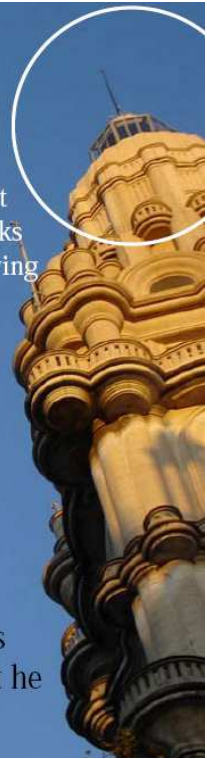}
\caption{The material representation of the angelic choruses of Dante's Paradise.}\label{fig0}\end{center}\end{figure} 

In these verses, the ``beauteous planet, that to love incites'' refers, of course, to planet Venus, which, in Dante's vision, dominated
with its light the eastern sky and covered with its luminosity the constellation Pisces (the Fish) that escorted it. Let us recall that
Dante had started his trip during Eastern, namely, when the Sun lies in Aries (the Ram), according to the traditional zodiac. Aries is the
sign bordering with Pisces. Hence, the fact that Venus --which is located always close to the Sun-- appears in the eastern sky tells us
that it is dawn. On the other hand, the fact that ``the beauteous planet'' appears on Pisces indicates precisely that this constellation
rose from the eastern horizon before Aries (and also before the Sun). These details, together with many others hinted by Dante during his
hasty voyage though the circles of Hell, tell us that the pilgrims are enjoying the dawn of resurrection (or Easter) day, during their
fourth day since the beginning of their trip.

\section{Final comments}

What does Dante mean by writing that he ``saw four stars ne'er seen before save by the primal people''? Could he be referring to the
Southern Cross? Can we think that Dante knew, {\it circa} 1300, that small constellation so characteristic of the southern hemisphere, and
that many of us employ regularly to find the celestial South Pole? Moreover, can we infer from this that Palanti wanted ``his'' Palace to
incorporate some feature allegorically related to the Cross? It may seem surprising that the poet, who never traveled southwards far enough
during his European wanderings, knew this group of southern stars not accessible from Florence. He might, however, have had news about
these stars in indirect ways, as Ptolemy's {\it Almagest} was known to him, apparently not directly, but through Alfraganus' {\it
  Compendium of Astronomy}. And in the {\it Almagest} these stars indeed appeared, not as a Cross, but as part of the constellation
Centaurus. 

However, it might also be that Dante ``invented'' these stars right away, with no relation to an existing star catalogue or travelers'
description whatsoever. To support this view, we have his explicit words ``four stars ne'er seen before save by the primal people'' which
might mean that only the inhabitants of the Garden of Eden, ``the primal people'' --Adam and Eve-- could have seen them, but not the rest
of ``their descendants''. Moreover, the poet just mentions the ``four stars'' but says nothing whether these formed part of an
astronomical asterism or constellation. 

Nonetheless, one can be pretty sure that Palanti wanted some celestial allegory to be attached to the highest --and more divine-- part of
the votive temple he built. The lighthouse, representing the nine angelical choruses surrounding God, with its powerful light that could
be seen from the opposite shore of the River Plate, had certainly some role in guiding the European pilgrims towards the South, towards
this promised land he might have thought Buenos Aires would be, and away from the Old Continent that would soon be devastated by a new
world war. Eventually, as an astronomical guide for pilgrims and travelers heading south, the Cross was the most appropriate ``sign'' in
the sky (Gangui 2009).

However, if Palanti really wanted those powerful twin beacons, the Salvo Palace (in Montevideo) and the Barolo Palace (on the Argentine
shore) to be related to Dante's entrance gate to the heavens, he probably should have thought more in Dante's ``glorious stars'',
specially in the twins Castor and Pollux (Pur. IV, 61) placed in ``Leda's lovely nest'', namely in Gemini, his birth sign, rather than in
the Southern Cross; the latter being a constellation Dante probably never saw with his own eyes and, may be, did not even cross his mind
either.

\end{document}